\documentclass[cameraready]{Interspeech}

\title{OmniCodec: Low Frame Rate Universal Audio Codec with Semantic–Acoustic Disentanglement}

\author[affiliation={1}]{Jingbin}{Hu}
\author[affiliation={1}]{Haoyu}{Zhang}
\author[affiliation={1}]{Dake}{Guo}
\author[affiliation={1}]{Qirui}{Zhan}
\author[affiliation={1}]{Wenhao}{Li}
\author[affiliation={1}]{Huakang}{Chen}
\author[affiliation={1}]{Guobin}{Ma}
\author[affiliation={1}]{Hanke}{Xie}
\author[affiliation={1}]{Chengyou}{Wang}
\author[affiliation={2}]{Pengyuan}{Xie}
\author[affiliation={2}]{Chuan}{Xie}
\author[affiliation={2}]{Qiang}{Zhang}
\author[affiliation={1}, correspondingauthor]{Lei}{Xie}

\address{
    $^1$ Audio, Speech and Language Processing Group (ASLP@NPU), \\
Northwestern Polytechnical University
\\
    $^2$ Shanghai Lingguang Zhaxian Technology 
}

\email{jingbin.hu@mail.nwpu.edu.cn, lxie@nwpu.edu.cn}


\usepackage{comment}
\usepackage[table]{xcolor}
\usepackage{makecell}
\usepackage{amsmath}
\definecolor{lightgreen}{RGB}{230,245,230}

\begin{document}

\maketitle

\begin{abstract}

Large Language Models (LLMs) have advanced audio generation through discrete representation learning. However, most existing neural codecs focus on speech and emphasize reconstruction fidelity, overlooking unified low frame rate modeling across diverse audio domains, including speech, music, and general sound. Moreover, high reconstruction quality does not necessarily yield semantically informative representations, limiting effectiveness in downstream generation tasks. We propose OmniCodec, a universal neural audio codec tailored for low frame rate. It adopts a hierarchical multi-codebook design with semantic–acoustic decoupling by leveraging the audio encoder of the pre-trained understanding model, along with a self-guidance strategy to improve codebook utilization and reconstruction. Compared with the Mimi codec, experiments show that OmniCodec achieves outstanding performance at the same bitrate, delivering superior reconstruction quality while also providing more semantically informative representations that benefit downstream generation tasks. Our model and code will be open-sourced\footnote{\url{https://github.com/ASLP-lab/OmniCodec}}. Our demo page is available \footnote{\url{https://hujingbin1.github.io/OmniCodec-Demo-Page/}}.
\end{abstract}

\section{Introduction}

In recent years, Large Language Models (LLMs) for audio generation have advanced rapidly~\cite{chen2025neural}, yet the core challenge remains the design of effective representations. Current approaches are generally based on discrete representations, which naturally align with the token-based space of language models. This alignment leads to more stable generation and better scalability for large-scale downstream generative tasks. Therefore, how to design a better Codec model as an Audio Tokenizer has become increasingly important.

In previous studies, the Codec model was mainly designed for audio compression and reconstruction tasks, such as SoundStream~\cite{zeghidour2021soundstream}, DAC~\cite{kumar2023high}, and Encodec~\cite{defossez2022high}, which have achieved high-quality reconstruction. However, due to the requirement for high fidelity, they usually need extremely high bit rates and frame rates, maintain a multi-codebook structure, and have a significant difference from semantic representations. All these factors make them unsuitable as targets for LLMs to predict.

With the advancement of technology, the single-codebook structure was highly valued due to its natural compatibility with language model modeling, such as BigCodec~\cite{xin2024bigcodec}, SingleCodec~\cite{li2024single}, TS3-Codec~\cite{wu2024ts3}, and FocalCodec~\cite{della2025focalcodec}. However, the information contained in the audio is very rich. A single-layer codebook is insufficient to represent the audio signal at low frame rate, and the high frame rate structure is not conducive to the data scaling training of LLM. Moreover, a Codec trained only based on reconstruction loss also has the common problem that semantic information is not structurally modeled.

\begin{figure*}[t]
  \centering
  \includegraphics[width=0.9\textwidth]{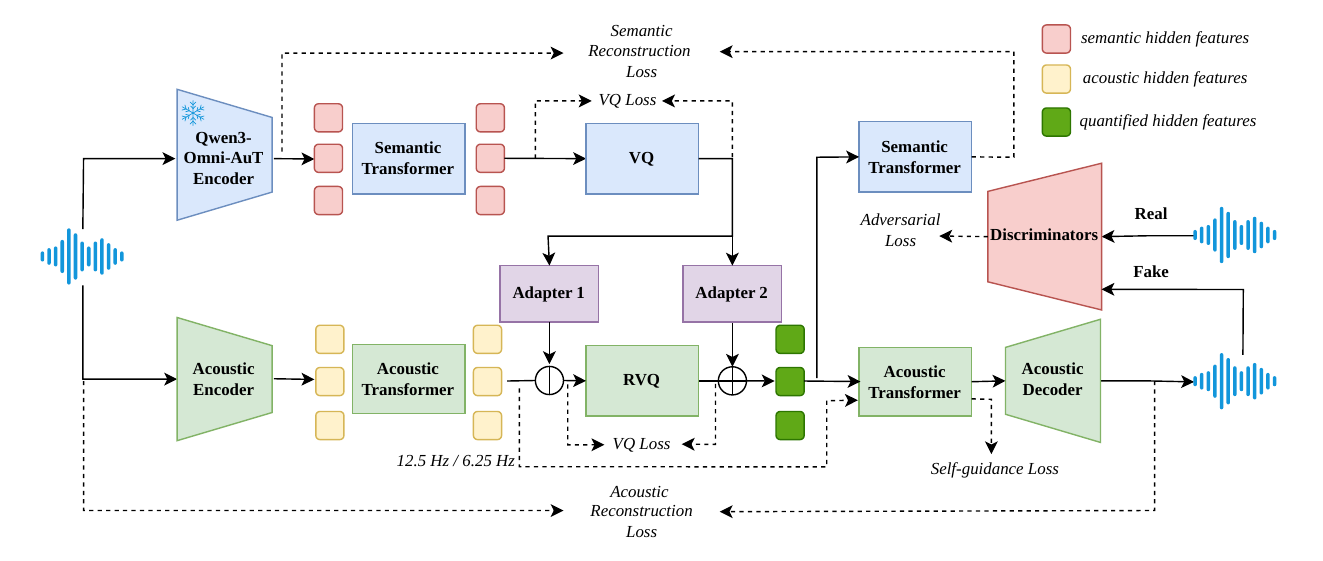}
  \caption{Overview of OmniCodec framework.}
  \label{fig: OmniCodec}
\end{figure*}

To address the issue of the absence of semantic information representation, some studies have attempted to use self-supervised learning models for distillation, facilitating the modeling of downstream tasks, such as SpeechTokenizer~\cite{zhang2023speechtokenizer}, Mimi codec~\cite{defossez2024moshi}, and Qwen-TTS-Tokenizer-12Hz~\cite{hu2026qwen3}. Some additional work attempts to use the representations extracted by the self-supervised learning model as the input for the semantic branch of the dual-stream Encoder, such as BiCodec~\cite{wang2025spark}, X-Codec~\cite{ye2025codec}, X-Codec2~\cite{ye2025llasa}, DualCodec~\cite{li2025dualcodec}, XY-Tokenizer~\cite{gong2025xy}, FireRedTTS-2-Tokenizer~\cite{xie2025fireredtts}, and SAC~\cite{chen2025sac}. Some radical proposals pursue extreme semanticity. They use the Automatic Speech Recognition (ASR) supervised loss for training the Codec model, such as S3Tokenizer~\cite{du2024cosyvoice} and GLM-4-Voice-Tokenizer~\cite{zeng2024glm}. These advancements have demonstrated the effectiveness of incorporating semantic information in enhancing the modeling capabilities of downstream LLMs. However, due to the limited capacity of a single codebook and the difficulty in achieving high-quality reconstruction at a low frame rate. These studies have shown that low frame rate and multi-codebook structures with semantic supervision are increasingly becoming the current development trend.

At the same time, general audio domains generation based on LLMs, which includes speech, music, and general sound, is becoming a strong demand. This demand has led to the development of discrete representations in the general audio domains. Early works, such as WavTokenizer~\cite{ji2024wavtokenizer} and SemantiCodec~\cite{liu2024semanticodec}, attempt to unify the modeling of audio in different domains. In recent work, UniCodec~\cite{jiang2025unicodec} attempts to enhance its semantic modeling capabilities through self-supervised learning methods. AUV~\cite{chen2025auv} has further advanced through cross-domain self-supervised feature distillation. Recently, there have also been some works to introduce LLMs at the bottleneck for speech recognition tasks and audio caption tasks, thereby introducing semantic supervision, such as the MiMo-Audio-Tokenizer~\cite{zhang2025mimo} and the MOSS-Audio-Tokenizer~\cite{gong2026moss}. By further scaling the model parameters and dataset size, they achieved extremely high reconstruction quality in general audio domains and low frame rate scenarios while retaining sufficient semantic information.

Although recent works have demonstrated the effectiveness of scaling in improving tokenizer performance, there remains a significant gap between academic research and industry resources. This gap makes it challenging to explore effective structural innovations. Additionally, large-scale tokenizers are not ideal for real-time speech interaction in streaming scenarios. With the further development of audio understanding models, the audio encoder used as the audio input for understanding models has been trained on vast datasets covering speech, music, and general sound, leading to robust semantic feature representations through supervised learning. This inspires us to explore using the audio encoder from understanding models as the semantic branch input, aiming to obtain general audio domain representations with semantic properties. By leveraging the data advantages from an open-source understanding model, we aim to achieve better semantic representations using the Qwen3-Omni-AuT-Encoder~\cite{xu2025qwen3}. 

To adapt to the trend of low frame rate, multiple codebooks,  general audio domains, and semantic decoupling, and also to integrate the audio encoder capabilities of the understanding model, we propose \textit{OmniCodec}. We present the first demonstration that an audio encoder trained using supervised learning from understanding models can replace unsupervised models like WavLM~\cite{chen2022wavlm} for use as semantic supervision in Codec models. And we have achieved the all-domain decoupling of speech, music, and general sound into semantic and acoustic representations. Moreover, we design a self-guidance mechanism~\cite{liself} to improve codebook utilization and stabilize training, leading to improved reconstruction quality. Furthermore, our model achieves two frame rates of 12.5 Hz and 6.25 Hz, and is designed with a purely causal receptive field, supporting fully streaming and fast inference.

Experiments have proved that our reconstruction indicators under the same bit rate condition are significantly superior to other competitive models. Moreover, in the downstream task, compared with the Mimi codec, we also achieved better results in the music and general sound domains. To facilitate future research, we open-source all models and code and provide a public demo showcasing generation examples across multiple audio domains.

\section{Method}

\begin{table*}[t]
\centering
\caption{Reconstruction evaluation on LibriSpeech test-clean, GTZAN testset, and a subset of AudioSet eval. 
The best, second-best, and third-best results are highlighted in bold, underline, and italic, respectively. 
Columns from left to right correspond to speech, music, and general sound evaluation. Light green rows indicate our models' performances. OmniCodec-F means OmniCodec-Flash. }
\label{tab: recon_all}
\small
\resizebox{\textwidth}{!}{
\begin{tabular}{lcc cccccccc}
\toprule
\textbf{Model}
& \textbf{\makecell{Codebook\\Size}}
& \textbf{\makecell{Bitrate\\(bps)}}
& \textbf{TPS}
& \textbf{PESQ-WB} $\uparrow$ 
& \textbf{PESQ-NB} $\uparrow$ 
& \textbf{STOI} $\uparrow$ 
& \textbf{Mel dis.} $\downarrow$
& \textbf{MCD} $\downarrow$
& \textbf{N-MOS} $\uparrow$
& \textbf{S-MOS} $\uparrow$ \\
\midrule

WavTokenizer & 4096 & 480 &40
&  1.88 / 1.14 / 1.35 & 2.44 / 1.51 / 1.77  & 0.87 / 0.49 / 0.36  & 0.97 / 1.16 / 1.10  & 4.79 / 4.47 / 5.08 & 3.02$\pm$0.11 & 3.11$\pm$0.13 \\

UniCodec & 16384 & 1050 & 75
&  2.65 / 1.33 / 1.59 & 3.14 / 1.85 / 2.15 & 0.92 / 0.64 / 0.49 & \underline{0.79} / 1.00 / 0.96  & 3.46 / 3.66 / 4.11 & 3.20$\pm$0.10 & 3.31$\pm$0.16  \\

AUV & 20480 & 716 & 50
& 2.40 / 1.29 / 1.54 & 3.05 / 1.91 / 2.13 & 0.91 / 0.60 / 0.47 & 0.89 / 1.26 / 1.26 & 3.60 / 3.94 / 4.37 & 3.15$\pm$0.17 & 3.23$\pm$0.08 \\

X-Codec-8L & 1024 & 4000 & $50\!\times\!8$
& \underline{2.96} / \textbf{1.95} / \textbf{2.07} & \underline{3.49} / \textbf{2.73} / \textbf{2.72} & \textit{0.93} / \underline{0.77} / \underline{0.65} & \underline{0.79} / \underline{0.87} / \textit{0.88} & \textit{3.35} / \textit{3.15} / 4.05 & \underline{3.56$\pm$0.07} & \underline{3.67$\pm$0.13} \\

Mimi codec-16L  & 2048 & 2200 & $12.5\!\times\!16$
& \textit{2.88} / \textit{1.56} / 1.71 & \textit{3.42} / \textit{2.12} / 2.39 & \underline{0.94} / \textit{0.74} / 0.60 & 1.12 / 1.18 / 1.14 & 3.93 / 3.56 / 4.06 & 3.42$\pm$0.09 & \textit{3.51$\pm$0.18} \\

\midrule

\rowcolor{lightgreen}
OmniCodec-32L & 2048 & 4400 & $12.5\!\times\!32$
& \textbf{3.02} / \underline{1.59} / \underline{1.89} & \textbf{3.62} / \underline{2.21} / \underline{2.60} & \textbf{0.96} / \textbf{0.78} / \textbf{0.66} & \textbf{0.75} / \textbf{0.86} / \textbf{0.82} & \textbf{2.54}/ \textbf{2.28} / \textbf{2.74} & \textbf{3.63$\pm$0.13} & \textbf{3.73$\pm$0.06} \\

\rowcolor{lightgreen}
OmniCodec-16L & 2048 & 2200 & $12.5\!\times\!16$
& 2.76 / 1.44 / \textit{1.75} & 3.41 / 2.02 / \textit{2.44} & \underline{0.94} / 0.73 / \textit{0.61} & \textit{0.81} / \textit{0.90} / \underline{0.85} & \underline{3.05} / \underline{2.68} / \underline{3.04} & \textit{3.45$\pm$0.05} & 3.50$\pm$0.09 \\

\rowcolor{lightgreen}
OmniCodec-8L & 2048 & 1100 & $12.5\!\times\!8$
& 2.14 / 1.31 / 1.55 & 2.78 / 1.75 / 2.16 & 0.91 / 0.66 / 0.52 & 0.90 / 0.97 / 0.90 & 3.78 / 3.30 / \textit{3.50} & 3.26$\pm$0.06 & 3.29$\pm$0.16 \\

\rowcolor{lightgreen}
OmniCodec-F-32L & 2048 & 2200 & $6.25\!\times\!32$
& 2.52 / 1.35 / 1.61 & 3.19 / 1.82 / 2.28 & 0.93 / 0.70 / 0.58 & 0.84 / 0.94 / 0.88 & 3.15 / 2.87 / 3.26 & 3.41$\pm$0.11 & 3.47$\pm$0.07 \\

\rowcolor{lightgreen}
OmniCodec-F-16L & 2048 & 1100 & $6.25\!\times\!16$
&  2.14 / 1.25 / 1.49 & 2.79 / 1.68 / 2.13 &  0.90 / 0.66 / 0.50 & 0.89 / 0.98 / 0.90 & 3.83 / 3.36 / 3.61 & 3.21$\pm$0.04 & 3.27$\pm$0.14 \\

\bottomrule
\end{tabular}
}

\end{table*}

\subsection{Universal neural audio codec}
As shown in Figure \ref{fig: OmniCodec}, our model is divided into two branches: semantic and acoustic. The encoder and decoder of acoustic use the SEANet with streaming convolution, and the Transformer uses a causal receptive field. Qwen3-Omni-AuT-Encoder is a pre-trained model that compresses the audio from 16kHz to 12.5Hz and represents it with a high-dimensional semantic feature layer. 

The semantic and acoustic branches discretize features respectively using vector quantization (VQ) and residual vector quantization (RVQ), and at the same time, through the Adapter, we perform the decoupling and recombination of the representation. For VQ and RVQ, we update the codebook using the moving exponential smoothing (EMA) method. For adversarial training, we employ three complementary discriminators: (1) A multi-scale STFT discriminator operating at multiple spectral resolutions, (2) A multi-scale, multi-period, and multi-band frequency-domain discriminators (MPD, MSD, and MRD) to enhance the frequency-domain consistency modeling of audio signal, and (3) A WavLM-based discriminator that provides high-level perceptual and semantic supervision. 

Our overall optimized target loss is as follows:
\begin{equation}
\begin{aligned}
L_{\text{total}} = &\ \lambda_{\text{ac\_recon}} L_{\text{ac\_recon}} + 
                   \lambda_{\text{se\_recon}} L_{\text{se\_recon}} + \\
                   &\lambda_{\text{commit}} L_{\text{commit}} + 
                   \lambda_{\text{self\_guidance}} L_{\text{self\_guidance}} + \\
                   &\lambda_{\text{dis}} L_{\text{dis}} + 
                   \lambda_{\text{gen}} L_{\text{gen}} + 
                   \lambda_{\text{fm}} L_{\text{fm}},
\end{aligned}
\end{equation}
The total optimization loss \( L_{\text{total}} \) includes the multiscale mel reconstruction loss \( L_{\text{ac\_recon}} \), the semantic representation reconstruction loss \( L_{\text{se\_recon}} \), the vector quantization commitment loss \( L_{\text{commit}} \), the self-guidance loss \( L_{\text{self\_guidance}} \), the adversarial losses \( L_{\text{dis}} \) and \( L_{\text{gen}} \), and the feature matching loss \( L_{\text{fm}} \). The weight for \( L_{\text{ac\_recon}} \) is 15.0, the weight for \( L_{\text{self\_guidance}} \) is 0.1, and the weights for the other terms are 1.0.

\subsection{Semantic and acoustic decoupling}

Qwen3-Omni-AuT-Encoder is an attention-encoder-decoder model, and it's trained on 20 million hours of supervised audio data, which has learned stronger and more general-purpose audio representations~\cite{xu2025qwen3}. It provides us with powerful semantic representations in general audio domains.

Moreover, inspired by DualCodec~\cite{li2025dualcodec}, we adopt the strategy of semantic and acoustic decoupling. We use the quantified semantic hidden features as the input for the adapter, which is a simple linear layer, instead of the predicted semantic features. In the acoustic branch, we first subtract the quantified semantic hidden features from the acoustic hidden features, and then add the quantized acoustic hidden features to the quantified semantic hidden features, thereby achieving decoupling.

\subsection{Self-guidance}

The core idea of self-guidance loss is to guide the decoder to produce similar outputs when processing both quantized tokens $z_q$ and continuous pre-quantized latents $z_e$. This loss guides the decoder to mimic the high-fidelity output that it would produce from continuous pre-quantized latents, thus improving its ability to handle quantization error and generating higher-quality reconstructions.

Formally, the self-guidance loss is defined as follows:
\begin{equation}
    L_{\text{self\_guidance}} = | \text{sg}(h_e) - h_q |_2^2,
\end{equation}

\begin{flushleft}

where sg(·) denotes the stop-gradient operation, $h_e$ and $h_q$ are hidden layer features after acoustic transformer, $h_e$ from the $z_e$ branch and $h_q$ from the $z_q$ branch. This loss term is added to form an end-to-end self-supervised training process~\cite{liself}. Our experiments have shown that this loss can be utilized to enhance the utilization rate of the codebook and the quality of the reconstruction.
\end{flushleft}

\begin{table*}[t]
\centering
\caption{Ablation study. Reconstruction evaluation on LibriSpeech test-clean and semantic evaluation trained on Emilia. Light green rows indicate our proposed models' performances.}
\label{tab: ablation}
\setlength{\tabcolsep}{2pt}
\resizebox{\textwidth}{!}{
\begin{tabular}{lcc cccccccc}
\toprule
\textbf{Ablation Study}
& \textbf{PESQ-WB} $\uparrow$ 
& \textbf{PESQ-NB} $\uparrow$ 
& \textbf{STOI} $\uparrow$ 
& \textbf{Mel dis.} $\downarrow$
& \textbf{MCD} $\downarrow$
& \textbf{PPL0} $\downarrow$
& \textbf{PPL mean} $\downarrow$ 
& \textbf{\makecell{Codebook\\Utilization}} $\uparrow$ 
& \textbf{N-MOS} $\uparrow$ 
& \textbf{S-MOS} $\uparrow$ 
\\
\midrule

\rowcolor{lightgreen}
OmniCodec-16L & 2.76 & 3.41 & 0.94 & 0.81  & 3.05 & 10.02 & 116.94 & 0.982 & 3.62$\pm$0.12 & 3.47$\pm$0.18 \\

\midrule
w/o Semantic branch 
 & 2.81 & 3.45 & 0.94 & 0.79 & 3.01 &  18.44 & 207.21 & 0.981 & 3.65$\pm$0.14 & 3.50$\pm$0.19  \\

w/o Self-guidance loss 
& 2.75 & 3.38 & 0.94 & 0.82  & 3.07 &  10.89 & 117.51 & 0.974 & 3.60$\pm$0.09 & 3.45$\pm$0.13  \\

w/o Adapter-1
& 2.61 & 3.29 & 0.94 & 0.76 & 3.05 &  11.13 & 119.80 & 0.980 & 3.59$\pm$0.07 & 3.43$\pm$0.10 \\

w/o music and general sound dataset
& 2.90 & 3.51 & 0.95 & 0.71 & 2.79 & 8.03 & 94.53 & 0.969 & 3.78$\pm$0.15 & 3.54$\pm$0.16  \\

\bottomrule
\end{tabular}
}
\end{table*}

\section{Experiments}
\subsection{Experimental setup}
\subsubsection{Datasets}
We train OmniCodec on approximately 160,000 hours of data. For speech, we use 95K-hour Emilia~\cite{he2024emilia} and LibriTTS~\cite{zen2019libritts}. For music, the in-house dataset is around 60K hours. For sound, we filter the Audio Set~\cite{gemmeke2017audio} with labels to form an 800-hour non-human sound set. The speech, music,
and sound reconstruction performances are evaluated on LibriSpeech test-clean~\cite{panayotov2015librispeech}, GTZAN testset~\cite{sturm2013gtzan}, and Audio Set eval.

\subsubsection{Pipeline}
All samples are resampled to 24 kHz. The samples are further downsampled by a factor of 1920 or 3840, yielding a token rate of 12.5 Hz or 6.25 Hz. Segments longer than 10 seconds will be truncated during training.For the Qwen3-Omni-AuT-Encoder, the input audio is resampled to 16 kHz. The audio is then processed by it, which downsamples the temporal resolution to 12.5 Hz and produces semantic hidden representations at this rate. We use a global batch size of 24, with gradient accumulation set to 2, across 4 A100 GPUs. The AdamW optimizer has a peak learning rate of 1e-4, where the scheduler linearly warmed up for 2.5K updates, cosine decayed for 500K updates. The model contains approximately 134M parameters. The SEANet backbone has a hidden dimension of 512, and hierarchical downsampling ratios of [8, 6, 5, 4] or [12, 8, 5, 4]. Semantic representations are discretized by a semantic vector quantizer with a codebook size of 2048 and an embedding dimension of 1024. Fine-grained acoustic details are modeled using a 31-stage residual vector quantizer, each with a codebook size of 2048 and 256-dimensional code vectors, with quantizer dropout enabled. A causal Transformer with 8 layers, 8 attention heads, model dimension 512, and feedforward size 2048.

\begin{table}[t]
\centering
\caption{Music and general sound domain reconstruction results. Audiobox Aesthetics scores include Content Enjoyment (CE), Content Usefulness (CU), Production Complexity (PC), and Production Quality (PQ).}
\label{tab: audiobox}
\setlength{\tabcolsep}{0.3pt}
\begin{tabular}{lcccc cccc}
\toprule
\textbf{Model}
& \multicolumn{4}{c}{\textbf{\makecell{GTZAN testset}}} 
& \multicolumn{4}{c}{\textbf{\makecell{AudioSet eval}}} \\
\cmidrule(lr){2-5} \cmidrule(lr){6-9}

& \textbf{CE} $\uparrow$ 
& \textbf{CU} $\uparrow$ 
& \textbf{PC} $\uparrow$ 
& \textbf{PQ} $\uparrow$ 
& \textbf{CE} $\uparrow$ 
& \textbf{CU} $\uparrow$ 
& \textbf{PC} $\uparrow$ 
& \textbf{PQ} $\uparrow$ \\

\midrule


Ground Truth & 7.44 & 7.41 & 6.40 & 7.52 & 4.03 & 5.62 & 3.71 & 6.10 \\
\midrule
Mimi codec-16L & 6.85 & 6.84 & 6.30 & 7.06 & 3.72 & 5.10 & \textbf{3.92} & 5.82 \\
OmniCodec-16L & \textbf{7.17} &\textbf{7.13} & \textbf{6.41} & \textbf{7.24} & \textbf{3.85} & \textbf{5.37} & 3.81 & \textbf{5.88} \\

\bottomrule
\end{tabular}
\end{table}

\begin{table}[t]
\centering
\caption{Semantic evaluation. We report LLM perplexity, which is PPL0 and PPL mean, trained on Emilia, the in-house dataset, and AudioSet to quantify semantic preservation on speech, music, and general sound domains.} 

\label{tab: semantic evaluation}
\setlength{\tabcolsep}{2pt}
\begin{tabular}{lccc}
\toprule
\textbf{Model}
& \textbf{Speech}
& \textbf{Music} 
& \textbf{Sound} \\

\midrule
DAC-8L & 19.84 / 237.26 & 8.54 / 113.09 &  4.60 / 67.96 \\
Mimi codec-8L & 8.73 / 102.70 & 4.43 / 51.08 &  3.74 / 42.33  \\
OmniCodec-8L & 10.02 / 116.94 & \textbf{4.14 / 48.92} &  \textbf{3.32 / 37.01} \\
OmniCodec-8L-FT & \textbf{8.42 / 97.94} & 4.78 / 59.23 &  3.81 / 46.56 \\

\bottomrule
\end{tabular}
\end{table}

\vspace{-3pt}
\subsubsection{Baselines}
\vspace{-3pt}

For general audio domains modeling, we compare our method with single-codebook approaches, including WavTokenizer~\cite{ji2024wavtokenizer}, UniCodec~\cite{jiang2025unicodec}, and AUV~\cite{chen2025auv}, as well as multi-codebook methods, including X-Codec~\cite{ye2025codec} and Mimi codec~\cite{defossez2024moshi}. 

\vspace{-3pt}
\subsubsection{Metrics}
\vspace{-3pt}

In reconstruction quality evaluation, for speech, we report Short Time Objective Intelligibility (STOI)~\cite{taal2010short}, Perceptual Evaluation
of Speech Quality (PESQ)~\cite{rix2001perceptual}, Mel distance (Mel dis.)~\cite{kubichek1993mel}, and Mel-cepstral Distance (MCD)~\cite{kubichek1993mel} in LibriSpeech test-clean. For music, we also report these in the GTZAN testset. For general sound, we report these in a subset of AudioSet eval. In the reconstruction evaluation experiment, 31 participants were invited to randomly select 30 reconstructed audio clips from each of the different domain test sets to score the N-MOS subjective indicators. For the S-MOS subjective indicator scoring, 30 randomly selected reconstructed audio clips from the LibriSpeech test-clean set were used. In the ablation experiment, the same participants, methods, and indicators were adopted in the speech test set. These indicators are reported with 95\% confidence intervals. In music and general sound, we also report the Audiobox Aesthetics scores in their testsets. In semantic evaluation, we report the perplexity (PPL) by training a 100M Qwen2-based LLM~\cite{yang2024qwen2technicalreport} trained on Emilia, the in-house dataset, and AudioSet. PPL0 represents the perplexity of the first-layer codebook in the validation set that is predicted, while PPL mean is the average perplexity of the eight-layer codebook in the validation set that is predicted~\cite{wang2025audiocodecbench}. Both of these indicators are used to measure the semantic performance of tokens and to validate the effect of downstream tasks.

\subsection{Reconstruction evaluation}

In the evaluation of the reconstruction task, as shown in Table \ref{tab: recon_all} and Table \ref{tab: audiobox}, under the same bit rate conditions, our model maintained a leading position in multiple reconstruction metrics across various domains. Notably, we significantly outperformed the Mimi codec in various indicators of Mel distance, MCD, and Audiobox scores, indicating that the model exhibits less distortion in the reconstructed spectrum and achieves better reconstruction perceptual quality in the music and general sound domains. At the same time, our model has a lower frame rate compared to the single-codebook model. Even for the 6.25Hz, 16-layer model, it outperforms the 75Hz UniCodec model in indicators such as STOI, Mel dis., and MCD, and has significantly lower frame rate, which demonstrates the effectiveness of our training strategy and model design. The subjective evaluation also demonstrated the excellent performance of the reconstructed audio from our model in terms of naturalness and speaker similarity.

\subsection{Semantic evaluation}

In the semantic evaluation, as shown in Table \ref{tab: semantic evaluation}, our model performed better in the music and general sound domains in terms of PPL compared to the Mimi codec, but performed worse in the speech domain. We think it is related to the structure of WavLM~\cite{chen2022wavlm}, which is based on the BERT~\cite{devlin2019bert} architecture and trained using a masked self-supervised learning approach, and is capable of learning more intricate phonetic details in speech. This unique structure allows it to capture subtle features that are particularly beneficial during speech domain distillation. As a result, WavLM provides a significant advantage, leading to higher PPL scores during evaluation in the speech domain. At the same time, considering the issue of data ratio, we attempted to only use the LibriTTS dataset to fine-tune OmniCodec for the speech domain, resulting in the OmniCodec-8L-FT model. We conducted a PPL experiment and found that the PPL in the speech domain was alleviated, but the PPL in the music and general sound domains increased. This indicates that the data ratio in different domains is also crucial. Compared to DAC, our model outperforms in both PPL0 and PPL mean, demonstrating the effectiveness of utilizing the audio encoder from a pre-trained understanding model, which provides robust semantic representations for the Codec model.

\subsection{Ablation study}

In the ablation experiment, as shown in Table \ref{tab: ablation}, we performed the ablation by removing the semantic branch and only conducting reconstruction, without using the self-guidance loss, removing the decoupled adapter1, and training only with speech data. We found that removing the semantic branch, the reconstruction metric would lead to an improvement, but it would cause a significant increase in the perplexity of the model in the downstream PPL experiment. Not using the self-guidance loss will result in a slight decrease in the reconstruction metrics, and under the same training conditions, the codebook utilization rate dropped from 0.982 to 0.974. Removing the decoupled adapter1 branch will cause a slight increase in PPL, while the reconstruction metrics will decrease, indicating the effectiveness of decoupling. Using only speech data achieves the best result on the speech test set, and the PPL also reaches the lowest value, demonstrating the importance of different domain data ratios. The subjective experiments further confirmed the issues reflected by the objective indicators.

\section{Conclusions}

In this work, we present a novel model that uses semantic knowledge from a pretrained Qwen3-Omni-AuT-Encoder into the first codebooks, while the remaining codebooks focus on acoustic reconstruction. Our model achieves all-domain decoupling of speech, music, and sound into semantic and acoustic representations, and introduces a self-guidance mechanism to improve codebook utilization and stabilize training, enhancing reconstruction quality. We show that using the audio encoder from the pre-trained understanding model provides robust semantic representations, boosting the Codec model's performance. However, challenges remain in speech decoupling, and future work will focus on adjusting data ratios, leveraging WavLM for joint distillation, and using larger datasets and training resources to further enhance semantic modeling.

\newpage
\section{Generative AI Use Disclosure}
In accordance with ISCA guidelines, the authors disclose the use of generative AI in the preparation of this manuscript. All intellectual contributions—including the core ideas, theoretical formulation, methodology design, experimental implementation, result analysis, and conclusions—were developed entirely by the authors without any involvement of generative AI tools. Generative AI was used solely for language polishing and stylistic refinement to improve the readability and fluency of the manuscript. The authors assume full responsibility for the intellectual content and scientific integrity of this work. No generative AI tool has been listed as a co-author, and all authors have reviewed and approved the final submission.

\bibliographystyle{IEEEtran}
\bibliography{mybib}

\end{document}